\documentclass[conference,letterpaper]{IEEEtran}

\usepackage[numbers,sort&compress]{natbib}

\makeatletter

\usepackage{etex}

\usepackage{zref-savepos}

\newcounter{mnote}

\def\xmarginnote{%
  \xymarginnote{\hskip -\marginparsep \hskip -\marginparwidth}}

\def\ymarginnote{%
  \xymarginnote{\hskip\columnwidth \hskip\marginparsep}}

\long\def\xymarginnote#1#2{%
\vadjust{#1%
\smash{\hbox{{%
        \hsize\marginparwidth
        \@parboxrestore
        \@marginparreset
\footnotesize #2}}}}}

\def\mnoteson{%
\gdef\mnote##1{\refstepcounter{mnote}\label{##1}%
  \zsavepos{##1}%
  \ifnum20432158>\number\zposx{##1}%
  \xmarginnote{{\color{blue}\bf $\langle$\arabic{mnote}$\rangle$}}%
  \else
  \ymarginnote{{\color{blue}\bf $\langle$\arabic{mnote}$\rangle$}}%
  \fi%
}
  }
\gdef\mnotesoff{\gdef\mnote##1{}}
\mnoteson
\mnotesoff

\usepackage{framed}
\usepackage{subcaption}
\usepackage{comment}
\usepackage[svgnames,dvipsnames]{xcolor}
\usepackage[all]{xy}
\usepackage{tikz}
\usetikzlibrary{positioning,matrix,through,calc,arrows,fit,shapes,decorations.pathreplacing,decorations.markings,}

\tikzstyle{block} = [draw,fill=blue!20,minimum size=2em]

\usepackage{qsymbols,amssymb,mathrsfs}
\usepackage{amsmath}
\usepackage[standard,thmmarks]{ntheorem}
\theoremstyle{plain}
\theoremsymbol{\ensuremath{_\vartriangleleft}}
\theorembodyfont{\itshape}
\theoremheaderfont{\normalfont\bfseries}
\theoremseparator{}

\theoremstyle{nonumberplain}
\theoremheaderfont{\scshape}
\theorembodyfont{\normalfont}
\theoremsymbol{\ensuremath{_\blacktriangleleft}}

\theoremnumbering{arabic}
\theoremstyle{plain}
\usepackage{latexsym}
\theoremsymbol{\ensuremath{_\Box}}
\theorembodyfont{\itshape}
\theoremheaderfont{\normalfont\bfseries}
\theoremseparator{}
\newtheorem{Conjecture}{Conjecture}

\theorembodyfont{\upshape}
\theoremprework{\bigskip\hrule}
\theorempostwork{\hrule\bigskip}

\usepackage[overload]{empheq} 

\let\iftwocolumn\if@twocolumn
\g@addto@macro\@twocolumntrue{\let\iftwocolumn\if@twocolumn}
\g@addto@macro\@twocolumnfalse{\let\iftwocolumn\if@twocolumn}

\mathtoolsset{showonlyrefs=false,showmanualtags}
\let\underbrace\LaTeXunderbrace 
\let\overbrace\LaTeXoverbrace
\renewcommand{\eqref}[1]{\textup{(\refeq{#1})}} 
\newtagform{brackets}[]{(}{)}   
\usetagform{brackets}

\usepackage[Smaller]{cancel}

\PassOptionsToPackage{breaklinks,letterpaper,hyperindex=true,backref=false,bookmarksnumbered,bookmarksopen,linktocpage,colorlinks,linkcolor=BrickRed,citecolor=OliveGreen,urlcolor=Blue,pdfstartview=FitH}{hyperref}
\usepackage{hyperref}

\usepackage{graphicx,psfrag}
\graphicspath{{figure/}{image/}} 

\usepackage{diagbox} 

\usepackage{algorithm2e}
\usepackage{listings} 
\lstdefinelanguage{Maple}{
  morekeywords={proc,module,end, for,from,to,by,while,in,do,od
    ,if,elif,else,then,fi ,use,try,catch,finally}, sensitive,
  morecomment=[l]\#,
  morestring=[b]",morestring=[b]`}[keywords,comments,strings]
\DeclareMathAlphabet{\mathpzc}{OT1}{pzc}{m}{it}
\usepackage{upgreek} 
\usepackage{dsfont}  

\def\multi@nostar#1#2{%
  \expandafter\def\csname multi#1\endcsname##1{%
    \if ##1.\let\next=\relax \else
    \def\next{\csname multi#1\endcsname}     
    \expandafter\newcommand\csname #1##1\endcsname{#2}
    \fi\next}}

\def\multi@star#1#2{%
  \expandafter\def\csname #1\endcsname##1{#2}
  \multi@nostar{#1}{#2}
}

\newcommand{\multi}{%
  \@ifstar \multi@star \multi@nostar}

\multi*{rm}{\mathrm{#1}}
\multi*{mc}{\mathcal{#1}}
\multi*{op}{\mathop {\operator@font #1}}
\multi*{ds}{\mathds{#1}}
\multi*{set}{\mathcal{#1}}
\multi*{rsfs}{\mathscr{#1}}
\multi*{pz}{\mathpzc{#1}}
\multi*{M}{\boldsymbol{#1}}
\multi*{R}{\mathsf{#1}}
\multi*{RM}{\M{\R{#1}}}
\multi*{bb}{\mathbb{#1}}
\multi*{td}{\tilde{#1}}
\multi*{tR}{\tilde{\mathsf{#1}}}
\multi*{trM}{\tilde{\M{\R{#1}}}}
\multi*{tset}{\tilde{\mathcal{#1}}}
\multi*{tM}{\tilde{\M{#1}}}
\multi*{baM}{\bar{\M{#1}}}
\multi*{ol}{\overline{#1}}

\multirm  ABCDEFGHIJKLMNOPQRSTUVWXYZabcdefghijklmnopqrstuvwxyz.
\multiol  ABCDEFGHIJKLMNOPQRSTUVWXYZabcdefghijklmnopqrstuvwxyz.
\multitR   ABCDEFGHIJKLMNOPQRSTUVWXYZabcdefghijklmnopqrstuvwxyz.
\multitd   ABCDEFGHIJKLMNOPQRSTUVWXYZabcdefghijklmnopqrstuvwxyz.
\multitset ABCDEFGHIJKLMNOPQRSTUVWXYZabcdefghijklmnopqrstuvwxyz.
\multitM   ABCDEFGHIJKLMNOPQRSTUVWXYZabcdefghijklmnopqrstuvwxyz.
\multibaM   ABCDEFGHIJKLMNOPQRSTUVWXYZabcdefghijklmnopqrstuvwxyz.
\multitrM   ABCDEFGHIJKLMNOPQRSTUVWXYZabcdefghijklmnopqrstuvwxyz.
\multimc   ABCDEFGHIJKLMNOPQRSTUVWXYZabcdefghijklmnopqrstuvwxyz.
\multiop   ABCDEFGHIJKLMNOPQRSTUVWXYZabcdefghijklmnopqrstuvwxyz.
\multids   ABCDEFGHIJKLMNOPQRSTUVWXYZabcdefghijklmnopqrstuvwxyz.
\multiset  ABCDEFGHIJKLMNOPQRSTUVWXYZabcdefghijklmnopqrstuvwxyz.
\multirsfs ABCDEFGHIJKLMNOPQRSTUVWXYZabcdefghijklmnopqrstuvwxyz.
\multipz   ABCDEFGHIJKLMNOPQRSTUVWXYZabcdefghijklmnopqrstuvwxyz.
\multiM    ABCDEFGHIJKLMNOPQRSTUVWXYZabcdefghijklmnopqrstuvwxyz.
\multiR    ABCDEFGHIJKL NO QR TUVWXYZabcd fghijklmnopqrstuvwxyz.
\multibb   ABCDEFGHIJKLMNOPQRSTUVWXYZabcdefghijklmnopqrstuvwxyz.
\multiRM   ABCDEFGHIJKLMNOPQRSTUVWXYZabcdefghijklmnopqrstuvwxyz.

\newcommand{\dotleq}{\buildrel \textstyle  .\over {\smash{\lower
      .2ex\hbox{\ensuremath\leqslant}}\vphantom{=}}}
\newcommand{\dotgeq}{\buildrel \textstyle  .\over {\smash{\lower
      .2ex\hbox{\ensuremath\geqslant}}\vphantom{=}}}

\DeclareMathOperator*{\argmax}{arg\,max}

\newcommand{\bM}{\begin{bmatrix}}
\newcommand{\eM}{\end{bmatrix}}
\newcommand{\bSM}{\left[\begin{smallmatrix}}
\newcommand{\eSM}{\end{smallmatrix}\right]}
\renewcommand*\env@matrix[1][*\c@MaxMatrixCols c]{%
  \hskip -\arraycolsep
  \let\@ifnextchar\new@ifnextchar
  \array{#1}}

\newqsymbol{`N}{\mathbb{N}}
\newqsymbol{`R}{\mathbb{R}}
\newqsymbol{`P}{\mathbb{P}}
\newqsymbol{`Z}{\mathbb{Z}}

\newqsymbol{`|}{\mid}
\newqsymbol{`8}{\infty}
\newqsymbol{`1}{\left}
\newqsymbol{`2}{\right}
\newqsymbol{`6}{\partial}
\newqsymbol{`0}{\emptyset}
\newqsymbol{`-}{\leftrightarrow}
\newqsymbol{`<}{\langle}
\newqsymbol{`>}{\rangle}

\DeclarePairedDelimiter\abs{\lvert}{\rvert}

\DeclarePairedDelimiter\Set{\{}{\}}
\newcommand{\imod}[1]{\allowbreak\mkern10mu({\operator@font mod}\,\,#1)}

\newcommand{\threecols}[3]{
\hbox to \textwidth{%
      \normalfont\rlap{\parbox[b]{\textwidth}{\raggedright#1\strut}}%
        \hss\parbox[b]{\textwidth}{\centering#2\strut}\hss
        \llap{\parbox[b]{\textwidth}{\raggedleft#3\strut}}%
    }
}

\newcommand{\reason}[2][\relax]{
  \ifthenelse{\equal{#1}{\relax}}{
    \left(\text{#2}\right)
  }{
    \left(\parbox{#1}{\raggedright #2}\right)
  }
}

\let\SavedDoubleVert\relax
\let\protect\relax
{\catcode`\|=\active
  \xdef\extendvert{\protect\expandafter\noexpand\csname extendvert \endcsname}
  \expandafter\gdef\csname extendvert \endcsname#1{\mskip-5mu \left.%
      \ifx\SavedDoubleVert\relax \let\SavedDoubleVert\|\fi
     \:{\let\|\SetDoubleVert
       \mathcode`\|32768\let|\SetVert
     #1}\:\right.\mskip-5mu}
}
\def\SetVert{\@ifnextchar|{\|\@gobble}
    {\egroup\;\mid@vertical\;\bgroup}}
\def\SetDoubleVert{\egroup\;\mid@dblvertical\;\bgroup}

\begingroup
 \edef\@tempa{\meaning\middle}
 \edef\@tempb{\string\middle}
\expandafter \endgroup \ifx\@tempa\@tempb
 \def\mid@vertical{\middle|}
 \def\mid@dblvertical{\middle\SavedDoubleVert}
\else
 \def\mid@vertical{\mskip1mu\vrule\mskip1mu}
 \def\mid@dblvertical{\mskip1mu\vrule\mskip2.5mu\vrule\mskip1mu}
\fi

\makeatother

\usepackage{fouridx}
\usepackage{framed}
\usetikzlibrary{positioning,matrix}

\usepackage{paralist}
\usepackage{enumerate}

\usepackage[normalem]{ulem}

\numberwithin{equation}{section}
\makeatletter
\@addtoreset{equation}{section}

\@addtoreset{Theorem}{section}

\@addtoreset{Lemma}{section}

\@addtoreset{Corollary}{section}

\@addtoreset{Example}{section}

\@addtoreset{Remark}{section}

\@addtoreset{Proposition}{section}

\@addtoreset{Definition}{section}

\@addtoreset{Claim}{section}

\@addtoreset{Subclaim}{Theorem}

\makeatother

{\endMakeFramed}

\newenvironment{ybox}{
	\setlength{\FrameSep}{1.5mm}
	\setlength{\FrameRule}{0mm}
  \MakeFramed {\FrameRestore}}%
{\endMakeFramed}

\newenvironment{gbox}{
	\setlength{\FrameSep}{1.5mm}
\setlength{\FrameRule}{0mm}
  \MakeFramed {\FrameRestore}}%
{\endMakeFramed}

{\endMakeFramed}

 {\endMakeFramed}

\usepackage{enumitem}

\usepackage{etoolbox}

\let\theparentequation\theequation
\patchcmd{\theparentequation}{equation}{parentequation}{}{}

\renewenvironment{subequations}[1][]{
	\refstepcounter{equation}%
	\setcounter{parentequation}{\value{equation}}
	\setcounter{equation}{0}
	\let\parentlabel\label
	\ifx\\#1\\\relax\else\label{#1}\fi
	\ignorespaces
}{%
	\setcounter{equation}{\value{parentequation}}
	\ignorespacesafterend
}

\newcommand*{\nextParentEquation}[1][]{
	\refstepcounter{parentequation}
	\setcounter{equation}{0}
	\ifx\\#1\\\relax\else\parentlabel{#1}\fi
}

\newcommand{\CS}{C_{\op{S}}}

\usepackage[outline]{contour}
\contourlength{1.2pt}

\title{Multiterminal Secret Key Agreement\\ at Asymptotically Zero Discussion Rate}

\author{Chung Chan, Manuj Mukherjee, Navin Kashyap and Qiaoqiao Zhou
	\thanks{C.\ Chan (email: chung.chan@cityu.edu.hk) is with the Department of Computer Science, City University of Hong Kong.}
        \thanks{Q.\ Zhou is with the Institute of Network Coding and the Department of Information Engineering, the Chinese University of Hong Kong.
	}
		\thanks{N.\ Kashyap (nkashyap@iisc.ac.in) and M.\ Mukherjee (manuj@iisc.ac.in) are with the Department of Electrical Communication Engineering, Indian Institute of Science, Bangalore 560012.}}

\begin{document}

\IEEEoverridecommandlockouts
\maketitle

\begin{abstract}
  In the multiterminal secret key agreement problem, a set of users want to discuss with each other until they share a common secret key independent of their discussion. We want to characterize the maximum secret key rate, called the secrecy capacity, asymptotically when the total discussion rate goes to zero. In the case of only two users, the capacity is equal to the G\'acs--K\"orner common information. However, when there are more than two users, the capacity is unknown. It is plausible that a multivariate extension of the G\'acs-K\"orner common information is the capacity, however, proving the converse is challenging. We resolved this for the hypergraphical sources and finite linear sources, and provide efficiently computable characterizations. We also give some ideas of extending the techniques to more general source models.
\end{abstract} 




\section{Introduction}
\label{sec:introduction}

We consider the multiterminal secret key agreement problem where a set of users want to agree on a common secret key after observing some private correlated sources and discussing in public at asymptotically zero rate. Following the work of \cite{bennett1988privacy}, which showed that public discussion helped agree on a secret key, the problem was formulated in the two-user case by \cite{maurer93,ahlswede93}. The model was later extended to the case with a helper in \cite{csiszar00} and the general multiterminal case in \cite{csiszar04} with arbitrary number of users and helpers. The goal is to characterize the maximum achievable secret key rate called the secrecy capacity.

The trade-off between the secrecy capacity and discussion rate was first studied in \cite{csiszar00}. However, the problem is difficult and only solvable or partially solvable in special cases, such as the case in \cite{csiszar00} with certain order of discussion, the two-user gaussian case in~\cite{watanabe10,watanabe11}, the high-rate regime where the secrecy capacity is maximized~\cite{tyagi13,LCV16,MKS16}, the multiterminal case in \cite{courtade16} with hypergraphical sources~\cite{chan10md} and linear discussion, and the multiterminal case in~\cite{chan16itw,chan17isit} with hypergraphical sources and the pairwise independent networks proposed in \cite{nitinawarat10,nitinawarat-ye10}.

We simplify the problem by considering the case with asymptotically zero discussion rate. Unlike the case with no discussion at all, some discussion is allowed as long as the rate is zero. While it is well-known that the secrecy capacity with no discussion is the G\'acs--K\"orner common information~\cite{gacs72} because the problem formulations are the same, the secrecy capacity at asymptotically zero discussion rate appears to be unknown in the general multiterminal case with interactive public discussion. To the best of our knowledge, other than the special discussion model in \cite{csiszar00}, the equivalence was known only in the two-user case for the general source model, following from the result of \cite{LCV16} (evaluated using the double Markov inequality as in \cite{tyagi13}). The proof techniques using Csisz\'ar sum inequality does not seem to extend to the multiterminal case.

In this work, we conjecture that the secrecy capacity with no discussion is equivalent to the case with asymptotically zero discussion in the general multiterminal case. We show that the conjecture holds for both the hypergraphical sources and finite linear sources, and obtain explicit characterizations of the corresponding secrecy capacities. In proving the results, we also strengthened an upper bound on the secrecy capacity in \cite{chan17isit} that uses the lamination technique in submodular function optimization. We also explain how the idea can be extended to more general source models to give non-trivial bounds.

\section{Problem formulation}
\label{sec:problem}

We are given a finite set $V:=\Set{1,\dots,m}$ of $m\geq 2$ users and a discrete memoryless
multiple source 
\begin{align*}
  \RZ_V:=(\RZ_i\mid i\in V)
\end{align*}
with the joint distribution denoted as $P_{\RZ_V}$ and a finite
alphabet set $Z_V:=\prod_{i\in V}Z_i$. (We will use sans serif font
for random variables and the normal font their alphabet set.) Each
user $i\in V$ can generate a private random variable $\RU_i$
independently, with $P_{\RU_V}=\prod_{i\in V}P_{\RU_i}$. Then, user
$i\in V$ observes privately an $n$-sequence $\RZ_i^n$ i.i.d.\
generated according to $\RZ_i$, with
$P_{\RZ_V^n|\RU_V}=P_{\RZ_V}^n$. 

The users can then discuss in public
interactively in multiple rounds. More precisely, at the $t$-th round,
for some $t\in \Set{1,2,\dots}$, some user $i_t\in V$ broadcast to
everyone in public the message
\begin{align*}
  \RF_t &:= f_t(\RF^{t-1},\tRZ_i),
\end{align*}
which is a function of the previous message $\RF^{t-1}:=(\RF_{\tau}\mid
\tau\leq t)$ and the private knowledge
$\tRZ_{i_t}:=(\RU_{i_t},\RZ_{i_t}^n)$ of user $i_t$. For convenience,
the entire sequence of public messages is denoted by
\begin{align*}
  \RF &:= (\RF_1,\RF_2,\dots).
\end{align*}

The users then identify and recover a secret key $\RK$, satisfying the
following recoverability and secrecy constraints: There exists some functions
$`f_i$ for $i\in V$ such that
\begin{align}
  \lim_{n\to`8} \Pr`1(\exists i\in V, \RK\neq `f_i(\RF,\tRZ_i)`2)&=0 \label{eq:recover}\\
  \limsup_{n\to `8} \frac1n`1[ \log \abs{K} - H(\RK|\RF)`2] &= 0.\label{eq:secrecy}
\end{align}
N.b., since we will focus on the converse proof techniques, weak secrecy is used to derive
stronger results. 

The secrecy capacity under the total discussion rate $R\geq 0$ is
defined as
\begin{align}
  \CS(R) &:= \liminf_{n\to `8} \frac1n \log \abs {K} \quad
           \text{such that} \label{eq:CS}\\
  &\limsup_{n\to`8} \frac1n \log \abs{F} \leq R \label{eq:R}
\end{align}
We are interested in characterizing $\CS(0)$, namely, the secrecy
capacity with asymptotically zero discussion rate.

\section{Preliminaries}
\label{sec:prelim}

Following from the result of G\'acs and K\"orner in \cite{gacs72}, a secret key rate
achievable without public discussion is as follows:
\begin{Proposition}
  \label{pro:>=JGK}
  $\CS(0)\geq J_{\op{GK}}(\RZ_V)$ where
  \begin{align}
    J_{\op{GK}}(\RZ_V) := \max\Set{ H(\RG) \mid H(\RG|\RZ_i)=0,
    \forall i\in V}\label{eq:JGK}
  \end{align}
  is called the (multivariate) G\'acs--K\"orner common information. 
\end{Proposition}
The optimal solution $\RG$ is called the maximum common function,
since it is a function of each $\RZ_i$, and its entropy is
maximized. It can be shown that every common function of $\RZ_i$'s is
a function of $\RG$. Although $\RG$ can be computed systematically
using the ergodic decomposition in \cite{gacs72}, the computation may take
exponential time.

The proof of the achievable result is quite straightforward because,
without any discussion, users can agree on $\RG^n$ perfectly with no
error. From $\RG^n$, a secret key of rate $H(\RG)$ can be extracted
by the usual compression technique. The
challenge is prove the converse and resolve the following conjecture:
\begin{Conjecture}
  \label{conj:CS=JGK}
  $\CS(0)=J_{\op{GK}}(\RZ_V)$.
\end{Conjecture}

To simplify the problem, we further consider the following source models.
\begin{Definition}[\mbox{\cite{chan10md}}]
  \label{def:hyp}
  The source $\RZ_V$ is said to be hypergraphical if, for
  all $i\in V$, $\RZ_i$ is equivalent to
  \begin{align}
    (\RX_e\mid e\in E, i\in `x(e)),
  \end{align}
  up to bijections,\footnote{$\RZ'_i$ is said to be a bijection of $\RZ_i$ iff $H(\RZ'_i|\RZ_i)=H(\RZ_i|\RZ_i')=0$.} where $E$ is the edge set and $`x:E\to 2^V`/\Set{`0}$ is called the
  edge function. The hypergraph $(V,E,`x)$ and the edge (random)
  variables $\RX_e$'s define the source.
\end{Definition}
A simple example of the hypergraphical source is as follows:
\begin{Example}
  \label{eg:hyp}
  Let $\RX_a$, $\RX_b$ and $\RX_c$ be uniformly random and independent
  bits. With $V:=\Set{1,2}$, define
  \begin{align*}
    \RZ_1:= (\RX_a,\RX_b,\RX_c), \quad \RZ_2:=(\RX_b,\RX_c),
    \,\text{ and }\,\RZ_3:=(\RX_a,\RX_c).
  \end{align*}
  This source is hypergraphical with $E=\Set{a,b,c}$,
  $`x(a)=\Set{1,2}$, $`x(b)=\Set{1,3}$ and $`x(c)=\Set{1,2,3}$.
\end{Example}

Another source model we will consider is:
\begin{Definition}[\mbox{\cite{chan10phd}}]
  \label{def:fls}
  The source $\RZ_V$ is said to be a finite linear source if, for all $i\in V$, $\RZ_i$ is equivalent to
  \begin{align}
    \RMx \MM_i,\label{eq:fls}
  \end{align}
  up to bijections, 
  where $\RMx$ is a uniformly random vectors with elements taking
  values from some finite field $\bbF_q$, and $\MM_i$ is a
  deterministic matrix with elements from $\bbF_q$. 
\end{Definition}
The following is an example of a finite linear source that is not hypergraphical.
\begin{Example}
  \label{eg:fls}
  Again with $\RX_a$, $\RX_b$ and $\RX_c$ being uniformly random and
  independent bits, define $V:=\Set{1,2,3}$,
  \begin{align*}
    \RZ_1 := \RX_a,\quad
    \RZ_2 := \RX_b,\,\text{ and }\,
    \RZ_3 :=\RX_a\oplus \RX_b,
  \end{align*}
  where $\oplus$ is the XOR operation. 
  This is a finite linear source because, with $\RMx:=\bM \RX_a & \RX_b \eM$,
  \begin{align*}
    \RZ_1 =\RMx \bM 1 \\ 0 \eM,\quad 
    \RZ_2 = \RMx \bM 0 \\ 1 \eM,\quad
    \RZ_3 = \RMx \bM 1\\ 1\eM, 
  \end{align*}
  and $\RMx$ is uniformly distributed over $\bbF_2^2$, where the matrix multiplications are over $\bbF_2$, and $\RMx$ is uniformly over $\bbF_2^2$. Note that $\RZ_i$'s are pairwise independent and so there is no edge variable with strictly positive entropy covering more than one node. There is no edge covering one node either, because each $\RZ_i$ is completely determined by other $\RZ_j$'s. However, $H(\RZ_V)=2>0$, and so it cannot be a hypergraphical source with no edge variable.
\end{Example}

\section{Main results}
\label{sec:results}

Conjecture~\ref{conj:CS=JGK} can be resolved in the affirmative for
both hypergraphical and finite linear
sources. The characterization of the capacity can also be evaluated more explicitly
and computed efficiently.

\begin{Theorem}
  \label{thm:hyp}
  For hypergraphical sources (in Definition~\ref{def:hyp}),
  $\CS(0)=J_{\op{GK}}(\RZ_V)$ with the optimal solution to
  \eqref{eq:JGK} being
  \begin{align}
    \RG = \RX_{\Set{e\in E\mid `x(e)=V}}, \label{eq:G:hyp}
  \end{align}
  namely the edge variables observed by every user.
\end{Theorem}
For the hypergraphical source defined in Example~\ref{eg:hyp}, we have
$\RG=\RX_c$ and so $\CS(0)=H(\RG)=H(\RX_c)=1$. To the best of our
knowledge, this simple result is not directly covered by any existing results.

\begin{Theorem}
  \label{thm:fls}
  For finite linear sources (in Definition~\ref{def:fls}),
  $\CS(0)=J_{\op{GK}}(\RZ_V)$ with the optimal solution to
  \eqref{eq:JGK} being
  \begin{align}
    \RG = \RMx \MM,
    \label{eq:G:fls}
  \end{align}
  where $\MM$ is a matrix whose column space is
  $`<\MM`>=\bigcap\nolimits_{i\in V} `<\MM_i`>$,
  namely the intersection of the column spaces of all $\MM_i$'s. $`<\MM`>$ is also the maximum common subspace $\argmax_{S} \Set{\dim S \mid S\subseteq `<M_i`> \,\forall i\in V}$.
\end{Theorem}
 
For the finite linear source defined in Example~\ref{eg:fls}, the
maximum common subspace of the column spaces of $\MM_i$'s is the
trivial vector space $\Set{\M0}$. A non-trivial example is given below.

\begin{Example}
  \label{eg:fls:2}
  Again with $\RX_a$, $\RX_b$ and $\RX_c$ being uniformly random and
  independent bits, define $V:=\Set{1,2}$,
  \begin{align*}
    \RZ_1 &:= (\RX_a, \RX_b, \RX_a\oplus \RX_b)\\
    \RZ_2 &:= (\RX_c,\RX_a\oplus \RX_b\oplus\RX_c).
  \end{align*}
  This is a finite linear source because, with $\RMx:=\bSM \RX_a,\RX_b,\RX_c\eSM$,
  \begin{align*}
    \RZ_1 =\RMx \overbrace{\bM 1 & 0 & 1\\ 0 & 1 & 1\\ 0 & 0 &
                                                               0\eM}^{\MM_1:=},\quad
    \RZ_2 = \RMx\overbrace{ \bM 0 & 1 \\ 0 & 1\\ 1 & 1 \eM}^{\MM_2:=},
  \end{align*}
  and $\RMx$ is uniformly distributed over $\bbF_2^2$.

  Before computing
  $\RG$ in \eqref{eq:G:fls}, notice that $\MM_1$ does not have full
  column rank because the last column is the sum of the first two. We
  may remove the last column and consider instead
  \begin{align}
    \RZ_1 &= \RMx \overbrace{\bM 1 & 0 \\ 0 & 1 \\ 0 & 0
                                                       \eM}^{\MM_1:=}\text{
                                                       and }
                                                       \RZ_2 = \RMx\overbrace{ \bM 0 & 1 \\ 0 & 1\\ 1 & 1 \eM}^{\MM_2:=}.\label{eq:fls:2}
  \end{align}
  To compute
  $`<\MM_1`>\cap `<\MM_2`>$, note that the null space of
  $\bSM \MM_1 &
  \MM_2 \eSM= `1[`1.\begin{smallmatrix} 1 & 0 \\ 0 & 1 \\ 0 & 0 \end{smallmatrix}`2|\begin{smallmatrix} 0 & 1\\ 0 & 1\\ 1 & 1\end{smallmatrix}`2]$ is
  spanned by $\bM \Mu \\ \Mv \eM $ with $\Mu=\Mv=\bSM 1 \\ 1
  \eSM$. Therefore, the matrix
                                      \begin{align}
                                        \MM:=\MM_1\Mu = -\MM_2\Mv = \bM1 & 1 & 0\eM^\intercal         \label{eq:fls:2:M}                               
                                      \end{align} 
spans the desired intersection $`<\MM_1`>\cap `<\MM_2`>$. Hence, $\RG
= \RMx\MM = \RX_a\oplus \RX_b$.
  \end{Example}

\section{Proofs}
\label{sec:proofs}

\subsection{Proof of Theorem~\ref{thm:hyp}}

To prove the result for hypergraphical sources, we will strengthen the
lamination bound in \cite[Theorem~4.3]{chan17isit} as follows:
\begin{Lemma}
  \label{lem:lam}
  For any hypergraphical sources and partition $\mcP$ of $V$ into at
  least two non-empty disjoint sets,
  \begin{subequations}
  \begin{align}
    &`a(\mcP) R \geq `1[1-`a(\mcP)`2]`1[\CS(R)-H(\RG)`2]
    \quad\text{where}\label{eq:lam}\\
    &`a(\mcP):=\frac{\max\limits_{e\in E: `x(e)\neq V} \raisebox{-.4em}{$\abs{\Set{C\in
      \mcP\mid C\cap `x(e)\neq `0}}-1$}}{\abs{\mcP}-1}\label{eq:`a}
  \end{align}
  \end{subequations}
  where $\RG:=\RX_{\Set{e\in E\mid `x(e)=V}}$ as defined in \eqref{eq:G:hyp}. 
\end{Lemma}
N.b., the original bound in \cite[(4.7)]{chan17isit} has neither the term $-H(\RG)$ nor the
condition $`x(e)\neq V$ in \eqref{eq:lam}.

To prove Theorem~\ref{thm:hyp} using the above lemma, it suffices to show that $`a(\mcP)\in
[0,1)$ for some partition $\mcP$, because then, \eqref{eq:lam} with $R=0$ implies $\CS(0)\leq
H(\RG)$. Since $\RG$ is a common function of $\RZ_i$'s, we have
$H(\RG)\leq J_{\op{GK}}(\RZ_V)$, which must be satisfied with equality
as desired by Proposition~\ref{pro:>=JGK}. Now, substitute into
\eqref{eq:`a} the partition $\Set {\Set {i}\mid i\in
  V}$ of $V$ into singletons:
\begin{align*}
  `a(\Set {\Set {i}\mid i\in V}) &= \frac{\max_{e\in E:`x(e)\neq V}
  \abs{`x(e)} -1}{\abs{V}-1}
\end{align*}
which is within $[0,1)$ as desired because  $`0\subsetneq`x(e)\subsetneq V$.

It remains to prove the above lemma.
\begin{Proof}[Lemma~\ref{lem:lam}]
  By the recoverability condition~\eqref{eq:recover}, for some
  $`d_n\to 0$, we have
  \begin{align*}
    n`d_n \geq \sum_{C\in \mcP} H(\RK|\RF,\tRZ_C) = \overbrace{\sum_{C\in \mcP}
    H(\RK,\RF|\tRZ_C)}^{`(1)} - \overbrace{\sum_{C\in \mcP} H(\RF|\tRZ_C)}^{`(2)}
  \end{align*}
  By \cite[Lemma~B.1]{csiszar04} for interactive discussion $\RF$,
  \begin{align*}
    `(2) &= (\abs{\mcP}-1)\sum_{C\in \mcP} \frac1{\abs{\mcP}-1} H(\RF|\tRZ_C)\\
    &\leq (\abs{\mcP}-1)H(\RF).
  \end{align*}
  
  To bound $`(1)$, let $\RZ'_V$ be the same hypergraphical source as $\RZ_V$ but with
  all edges $e\in E$ such that $`x(e)=V$ removed. For convenience,
  write $\bar\RZ_i$ for $({\RZ'}^n_i,\RU_i)$, just like $\tRZ_i$ for
  $(\RZ_i^n,\RU_i)$.
  Since $\RG$ is determined by $\RZ_i$ for any $i\in V$,
  \begin{align*}
    `(1) &= \sum_{C\in \mcP} H(\RK,\RF|\bar{\RZ}_C,\RG^n)\\
    &\geq \underbrace{`1[\sum_{C\in \mcP} 1 - \max_{e\in E: `x(e)\neq V}
      \sum_{C\in \mcP: `x(e)\cap C\neq `0} 1`2]}_{`(3)} H(\RK,\RF|\RG^n),
  \end{align*}
  where the last inequality is by the lamination technique. (See
  \cite[Proposition~B.1]{chan17isit-arxiv} and its application in
  \cite[(B.10)]{chan17isit-arxiv}.)
  \begin{align*}
    `(3) &= \abs{\mcP} - \max_{e\in E:`x(e)\neq V} \abs{\Set {C\in \mcP\mid `x(e)\cap C\neq `0}} \\
         &= (\abs{\mcP}-1)  [1-`a(\mcP)].
  \end{align*}
  Altogether, we have
  \begin{align*}
    n`d_n &\geq (\abs {\mcP}-1)  [1-`a(\mcP)] H(\RK,\RF|\RG^n) - (\abs{\mcP}-1)H(\RF)\\
    \frac{`d_n}{\abs{\mcP}-1} &\geq [1-`a(\mcP)] \frac{H(\RK|\RF)-H(\RG^n)}n -`a(\mcP) \frac{H(\RF)}n.
  \end{align*}
  Assuming the secret key agreement scheme achieves $\CS(R)$, the above inequality 
  implies \eqref{eq:lam} as desired because $\frac{H(\RG^n)}n=H(\RG)$ by independence,
  \begin{align*}
    \liminf_{n\to `8} \frac{H(\RK|\RF)}{n} \geq \liminf_{n\to `8} \frac1n \log \abs{K} = \CS(R)
  \end{align*}
  by the secrecy constraint \eqref{eq:secrecy} and the definition of the capacity~\eqref{eq:CS}, and
  \begin{align*}
    \limsup_{n\to `8} \frac{H(\RF)}n \leq \limsup_{n\to `8} \frac{\log\abs{F}}n \leq R
  \end{align*}
  by \eqref{eq:R}.
\end{Proof}

\subsection{Proof of Theorem~\ref{thm:fls}}

To prove the result for finite linear sources, we first show the base
case with two users, i.e., $V=\Set{1,2}$, and then extend it to the
more general case with multiple users. The base case follows immediately from that of hypergraphical sources because of the following observation, the proof of which will be given later in this section:
\begin{Lemma}
  \label{lem:fls2hyp}
  A finite linear source involving $\abs{V}=2$ users is hypergraphical.
\end{Lemma}
Unfortunately, the above result does not extend to $\abs{V}>2$. A
counter-example is in
Example~\eqref{eg:fls}, which gives a finite linear source that is not
hypergraphical. To prove the desired Theorem~\ref{thm:fls} with the
above lemma, we will use a more contrived argument below. 

First of all, similar to the proof of Lemma~\ref{lem:lam}, the recoverability constraint~\eqref{eq:recover} implies that, for some $`d_n\to 0$,
\begin{align*}
  n`d_n &\geq \underbrace{\sum_{i\in V} H(\RK,\RF|\tRZ_i)}_{`(1)} - \underbrace{\sum_{i\in V}H(\RF|\tRZ_i)}_{`(2)}
\end{align*}
where $`(2)\leq (\abs {V}-1) H(\RF)$ by \cite[Lemma~B.1]{csiszar04}. We will show by induction that
\begin{align}
  `(1) \geq H(\RK,\RF|\RG^n) \label{eq:*1}
\end{align}
where $\RG:=\RMx\MM$ as defined in \eqref{eq:G:fls} with $`<\MM`>$ being the maximum common subspace of the column spaces of $\MM_i$'s.
It follows that
\begin{align*}
  `d_n \geq \frac{H(\RK|\RF)-H(\RG^n)}n - (\abs {V}-1) \frac{H(\RF)}n
\end{align*}
which implies that $0\geq \CS(0)-H(\RG)$ with $R=0$ by the secrecy
constraint~\eqref{eq:secrecy}, definition~\eqref{eq:CS} of the
capacity and the discussion rate constraint~\eqref{eq:R}. The
inequality must be satisfied with equality because $\RG$ is a common
function of $\RZ_i$'s and so $H(\RG)\leq J_{\op{GK}}(\RZ_V)$ as desired.

To prove \eqref{eq:*1} by induction. For the base case $V=\Set{1,2}$,
\begin{align}
  &\kern-2em H(\RK,\RF|\tRZ_1) + H(\RK,\RF|\tRZ_2)\notag\\
  &=  H(\RK,\RF|\tRZ_1,\RG^n) + H(\RK,\RF|\tRZ_2,\RG^n)\notag\\
                                        &\geq H(\RK,\RF| \tRZ_1,\tRZ_2,\RG^n) + H(\RK,\RF|\RG^n)\label{eq:*}\\
                                        &\geq H(\RK,\RF|\RG^n)\notag
\end{align}
where the first equality is because $\RG$ is a common function of $\RZ_i$'s, the second inequality follows again from the lamination technique in~\cite[Proposition~B.1]{chan17isit-arxiv} since $(\RZ_1,\RZ_2)$ is hypergraphical by Lemma~\ref{lem:fls2hyp}. For the induction, consider $\abs{V}>2$ and any $j\in V$. Assume as an inductive hypothesis that
\begin{align}
  \sum_{i\in V`/\Set{j}} H(\RK,\RF|\tRZ_i) \geq H(\RK,\RF|\tRG^n) \label{eq:IH}
\end{align}
where $\tRG = \RMx \tMM$ and $`<\tMM`>=\bigcap_{i\in V`/\Set{j}}
`<\MM_i`>$. Then,
\begin{align*}
  \sum_{i\in V} H(\RK,\RF|\tRZ_i) &= \sum_{i\in V`/\Set{j}} H(\RK,\RF|\tRZ_i) + H(\RK,\RF|\tRZ_j)\\
                                  &\geq H(\RK,\RF|\tRG^n)+H(\RK,\RF|\tRZ_j,\RG^n)\\
  &\geq H(\RK,\RF|\tRG^n,\RG^n) + H(\RK,\RF|\RG^n)
\end{align*}
where the first inequality is by the inductive
hypothesis~\eqref{eq:IH} and the fact that $\RG$ is a common function
of all $\RZ_i$'s. The last inequality is again by the lamination
technique in~\cite[Proposition~B.1]{chan17isit-arxiv} because
$\RZ_1':=\tRG$ and $\RZ_2':=(\RZ_j,\RG)$ defines a finite linear
source $(\RZ_1',\RZ_2')$, which is also hypergraphical by
Lemma~\ref{lem:fls2hyp}. It remains to prove this lemma.

\begin{Proof}[Lemma~\ref{lem:fls2hyp}]
  By \eqref{eq:fls}, write $\RZ_1=\RMx \MM_1$ and $\RZ_2=\RMx \MM_2$, for some uniformly random $\RMx$ with elements from a finite field $\bbF_q$. Without loss of generality, we can choose $\MM_1$ and $\MM_2$ such that they both have full column ranks. This is because, if the column rank of $\MM_i$ is not full, any column of $\MM_i$ linearly dependent on others columns correspond to redundant observations that can be removed.

  Let $\MM$ be the matrix such that $`<\MM`>=`<\MM_1`>\cap `<\MM_2`>$ as in \eqref{eq:G:fls}. Without loss of generality, suppose
  \begin{align}
    \begin{split}
    \MM_1 &= \bM \MM & \MN_{1} \eM\\
    \MM_2 &= \bM \MM & \MN_{2} \eM
  \end{split}
                       \label{eq:fls2hyp:M}
  \end{align}
  for some matrices $\MN_1$. This is possible by some invertible transformations of the rows of $\MM_i$'s (post-multiplying an invertible matrix), because $`<\MM`>$ is a common subspace of 
  the column spaces of $\MM_1$ and $\MM_1$.

  It follows that
  \begin{align}
    \MT &:= \bM \MM &\MN_1 &\MN_2 \eM\label{eq:fls2hyp:T}
  \end{align}
  must have full column rank. Suppose to the contrary that $\MT$ does not have full column rank, i.e., $\bM \MM &\MN_1 &\MN_2 \eM \bM \Mu^\intercal &\Mv^\intercal &\Mw^\intercal \eM^\intercal=0$ for some row non-zero row vector $\bM \Mu^\intercal &\Mv^\intercal &\Mw^\intercal \eM^\intercal$. Then, $\Mv$ is non-zero because, otherwise, $\bM \Mu^\intercal &\Mw^\intercal \eM^\intercal$ is non-zero but $\MM_2\bM \Mu^\intercal &\Mw^\intercal \eM^\intercal=0$, contradicting the assumption that $\MM_2$ has full column rank. Similarly, $\Mw$ is non-zero. Hence, we can write $\MN_2\Mw=-\MM_1\bM \Mu^\intercal &\Mv^\intercal \eM^\intercal$, which is therefore in $`<\MM_1`>\cap `<\MM_2`>$ and therefore $`<\MM`>$, contradicting the assumption that $\MM_2$ has full column rank.

  To show that $(\RZ_1,\RZ_2)$ is hypergraphical, write
  \begin{align*}
    \RZ_1 = \overbrace{\RMx \MT}^{\RMx':=} \overbrace{\bM \MI & \M0
    \\ \M0 & \MI \\ \M0 & \M0 \eM}^{\MM_1':=},\,\text{ and }\, \RZ_2 = \underbrace{\RMx \MT}_{=\RMx'} \overbrace{\bM \MI & \M0 \\ \M0 & \M0 \\ \M0 & \MI \eM}^{\MM_2':=}
  \end{align*}
  where $\MI$ denotes the identity matrix.
  The above equalities can be easily verified by substituting the value of 
  $\MT$ in~\eqref{eq:fls2hyp:T} to give \eqref{eq:fls2hyp:M}. 
  Since $\MT$ has full column rank, the elements of $\RMx'$ are independent and uniformly random over $\bbF_2$. Furthermore, since every column of $\MM_1'$ and $\MM_2'$ contains only one non-zero entry, the source $(\RZ_1,\RZ_2)$ is hypergraphical.
\end{Proof}

We will illustrate the above proof using
Example~\ref{eg:fls:2}. Recall the source written in
\eqref{eq:fls:2} in terms of the matrices $\MM_1$ and $\MM_2$. Recall
also the matrix $\MM$ defined in \eqref{eq:fls:2:M} for \eqref{eq:G:fls} that spans the intersection of the column spaces
of $\MM_1$ and $\MM_2$. The orthogonal complements of $`<\MM`>$ in
$`<\MM_1`>$ and $`<\MM_2`>$ are spanned respectively by
\begin{align*}
  \MN_1:=\bM 0 & 1 & 0\eM^\intercal\,\text{ and }\,\MN_1:=\bM 0 & 0 & 1\eM^\intercal
\end{align*}
and so, as in \eqref{eq:fls2hyp:M}, we can equivalently consider
\begin{align*}
  \RZ_1 = \RMx \bSM \MM & \MN_1\eSM = \RMx \overbrace{\bM 1 & 0\\ 1 &
                                                                    1\\ 0 & 0\eM}^{\MM_1:=},
  \RZ_2 = \RMx \bSM \MM & \MN_2\eSM = \RMx \overbrace{\bM 1 & 0\\ 1 &
                                                                    0\\ 0 & 1\eM}^{\MM_2:=}.
\end{align*}
With $\MT$ defined in \eqref{eq:fls2hyp:T}, and
\begin{align*}
  \RMx' :=\RMx\MT&=\bM \RX_a & \RX_b & \RX_c \eM \bM 1 & 0 & 0\\ 1 & 1 & 0\\ 0 & 0 & 1\eM\\
        &=\bM\RX_a\oplus \RX_b, \RX_b, \RX_c\eM,\\[-1em]
        &\kern1.5em \underbrace{\hphantom{\RX_a\oplus \RX_b}}_{\RX_{a'}:=}
\end{align*}
we have $\RMx'$ uniformly distributed over $\bbF_2^3$,
\begin{align*}
  \RZ_1 = \RMx' \bM 1 & 0 \\ 0 & 1\\ 0 & 0\eM,\,\text{ and }\,
                                          \RZ_2=\RMx' \bM 1 & 0\\ 0 &
                                                                      0\\
  0 & 1\eM.
\end{align*}
Hence, $(\RZ_1,\RZ_2)$ is hypergraphical (see
Definition~\ref{def:hyp}) with 
$E=\Set{a',b,c}$, $`x(a')=\Set{1,2}$, $`x(b)=\Set{1}$ and $`x(c)=\Set{2}$.

\section{Extensions to more general sources}
\label{sec:extension}

In this work, we showed for hypergraphical and finite linear sources that the secrecy capacity at asymptotically
zero discussion rate is given by the multivariate G\'acs--K\"orner
common information. The main property for proving the results is the 
lamination technique in \cite{chan17isit} for hypergraphical
sources. In the case with two users $V=\Set{1,2}$, it simplifies to \eqref{eq:*}:
\begin{align*}
  H(\RF,\RK|\tRZ_1,\RG^n)&+  H(\RF,\RK|\tRZ_2,\RG^n)\\
 &\geq   H(\RF,\RK|\tRZ_1,\tRZ_2,\RG^n)+  H(\RF,\RK|\RG^n)
\end{align*}
where $\RG$ is the maximum common function of $\RZ_1$ and $\RZ_2$. The
proof for finite linear source also boils down to this case, by
noticing that a finite linear source for two users is a
hypergraphical source. For more general source models, however, the
above inequality does not hold, and so the techniques considered does
not directly extend.

It is easy to show, however, that the above inequality still holds for
the general sources if $\RG$ is replaced by a random variable $\RW$
that satisfies the Markov chain $\RZ_1"-"\RW"-"\RZ_2$. In particular,
$\RW$ can be the Wyner common information~\cite{wyner75} between $\RZ_1$ and
$\RZ_2$. This allows us
to derive upper bounds on $\CS(0)$ for general sources such as
\begin{align*}
  \CS(0) \leq H(\RW_m)
\end{align*}
for any $\RW_V$ with $\RW_1=\RZ_1$ and the Markov chains
$\RZ_{j+1}"-"\RW_{j+1}"-"\RW_j$ for all $j>1$. For
hypergraphical and finite linear sources, it can be shown that the
tightest bound is given by 
the choice of $\RW_j$ being the Wyner common information
between $\RZ_{j+1}$ and $\RW_j$. Since the Wyner common information is
the same as the G\'acs--K\"orner common information for two-user
hypergraphical or finite linear sources, the above bound is precisely
$J_{\op{GK}}(\RZ_V)$. In general, however, Wyner common information
may not be equal to the G\'acs--K\"orner common information, so the
bound may not be tight. There are also possible improvements to the
bound, by considering different ordering of elements in $V$, and impose a
Markov tree instead of a chain. Proving conjecture~\ref{conj:CS=JGK}
for general sources remains
an interesting open problem.


\bibliographystyle{IEEEtran}
\bibliography{IEEEabrv,ref}

\end{document}